\documentclass[conference, a4paper]{IEEEtran}
\IEEEoverridecommandlockouts
\usepackage{cite}
\usepackage{amsmath,amssymb,amsfonts}
\usepackage{algorithm}
\usepackage{algorithmic}
\usepackage{graphicx}
\usepackage{textcomp}
\usepackage{xcolor}
\usepackage{booktabs}
\usepackage{subfigure}
\usepackage{tikz}

\usetikzlibrary{decorations.pathreplacing}
\usetikzlibrary{decorations.pathreplacing,calligraphy}
\def\BibTeX{{\rm B\kern-.05em{\sc i\kern-.025em b}\kern-.08em
    T\kern-.1667em\lower.7ex\hbox{E}\kern-.125emX}}
\begin{document}

\title{Influence Maximization on Dynamic Social Networks with Conjugate Learning Automata}

\author{
\IEEEauthorblockN{Chong Di*, Fangqi Li*, Shenghong Li}
\IEEEauthorblockA{\textit{School of Electronic Information and Electrical Engineering,} \\
\textit{Shanghai Jiao Tong University}\\
Shanghai, China \\
*Equal Contribution \\
\IEEEauthorblockN{$\{$dichong95, solour\_lfq, shli$\}$@sjtu.edu.cn}}
}

\maketitle

\begin{abstract}
Selecting the optimal subset from all vertices as seeds to maximize the influence in a social network has been a task of interest. Various methods have been proposed to select the optimal vertices in a static network, however, they are challenged by the dynamics, i.e. the time-dependent variation of the social network structure.
Such dynamics hinder the paradigm for static networks and leaves a seemingly unbridgeable gap between algorithms of influence maximization on static networks and those on dynamic ones.

In this paper, we extend our previous work and demonstrate that conjugate learning automata (an elementary variant of reinforcement learning) that have been successfully applied to maximize influence on static networks can be applied to dynamic networks as well. The network dynamics is measured by the variation of the influence range and absorbed into the learning procedure. Our proposal delicately formulates the effect of network dynamics: the more the influence range varies, the more likely the seeds are to be learned from scratch. Under this assumption, the continuity of the network variation is fully taken advantage of. Experimental results on both synthetic and real-world networks verify the privileges of our proposal against alternative methods.
\end{abstract}


\section{Introduction}
\label{section:1}
The online social network has undergone diversified studies concerning its community structure, swarm behavior, etc. Among them, the task of influence maximization (IM) is of particular interest and significance \cite{ref:IM}. In IM, a social network is formulated as a graph $\mathcal{G}=(\mathcal{V},\mathcal{E})$, where $\mathcal{V}=\left\{v_{1},v_{2},\cdots,v_{N}\right\}$ denotes $N$ participants of this network, while their interconnections are embedded in $\mathcal{E}$.  IM aims to locate the optimal subset $\mathcal{S}^{*}\subset \mathcal{V}$ with $K$ participants as seeds such that information propagation from them can affect as many participants as possible. Current mainstream solutions to IM are greedy algorithms, topology-based methods and heuristic ones \cite{ref:IM}. It is remarkable that the solutions of IM can be applied to various scenarios with slight modification, examples including \cite{ref:App1}\cite{ref:App2}.


\begin{figure}[htbp]
\centering
\includegraphics[width = 0.48\textwidth]{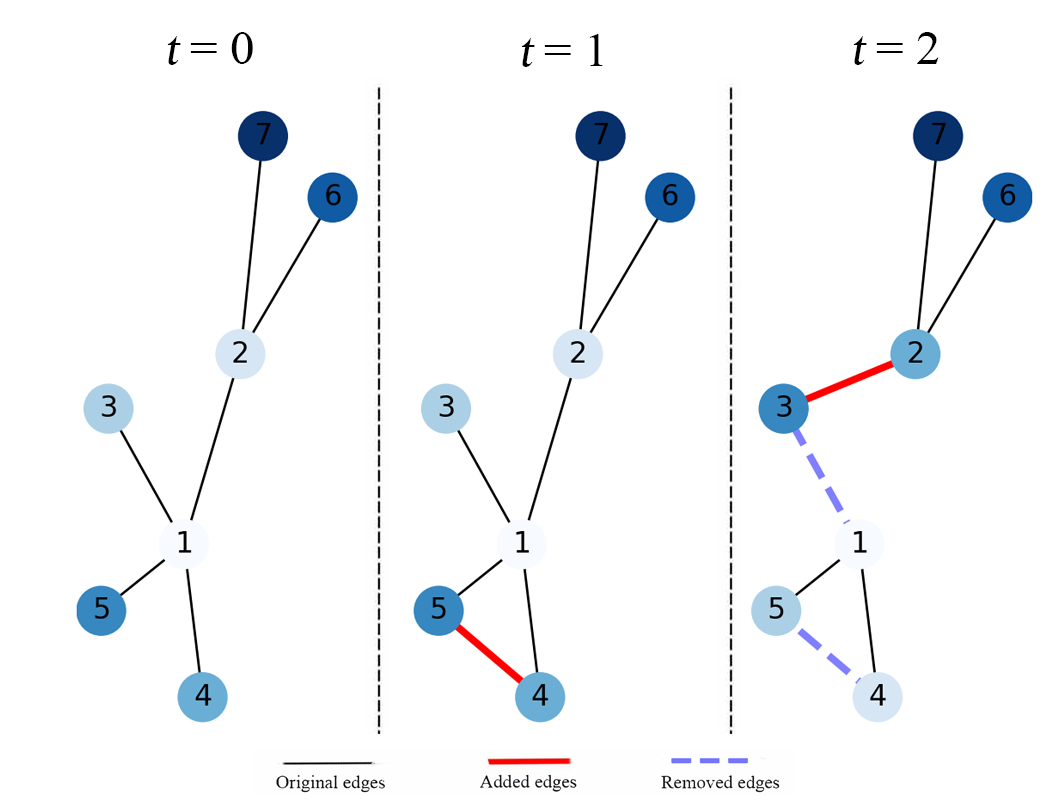}
\caption{The toy examples with $N=7$, $K=1$. The optimal seed is $v_{1},v_{1},v_{2}$ for $t = 0,1,2$.}
\label{figure:2}
\end{figure}

However, the structure of an online social network could vary with time, i.e. both $\mathcal{V}$ and $\mathcal{E}$ might change. Figure. \ref{figure:2} illustrates an example of dynamic network. The challenges derived from network dynamics split into two aspects: On one hand, it is hard to evaluate the influence of a topological change in a complex network, so it is unclear how the optimality of seeds from the previous snapshot can be preserved. For example, in Figure. \ref{figure:2}, when $t=0,1$ the optimal seed is $v_{1}$, however, at $t=2$, the optimal seed becomes $v_{2}$. On the other hand, if we abandon the previous knowledge and run an IM algorithm for any new snapshot independently, then it has to be very efficient, otherwise, we might fail to catch up with the update of the network structure. To address the two challenges, the evolved version of an IM algorithm on dynamic social networks should: (1). Not only consider local topology. (2). Take advantage of the historical information to cut down redundancy. There have been some trial solutions to IM on a dynamic social network by probing \cite{ref:dy1}\cite{ref:dy2} or formulating it as a bandit task \cite{ref:dy3}, etc.

Recently some researches applied Learning Automaton (LA), an elementary paradigm in reinforcement learning, to find a solution to IM \cite{ref:IMLA}. By modifying the traditional LA, Conjugate Learning Automata (CLA) has been proposed to evade potential adverse pitfalls that impair the performance of greedy algorithms and is comparatively efficient \cite{ref:IMCLA}. In CLA, $K$ individual LAs cooperate to find the optimal set of seeds through a learning procedure in which each LA is responsible for finding one candidate seed.

In this paper, we extend CLA to adapt to dynamic networks and propose Dynamic Conjugate Learning Automata (DyCLA). Roughly speaking, in DyCLA, each individual LA chooses a candidate seed if its learning process converges. Once the variation of the network structure takes place, the learning processes of individual LAs are rewound and a forked learning process is rehearsed in the new network. If the influence range only changes slightly, then the rewinding is also slight so DyCLA can recall previous knowledge and quickly converge to a new optimum. If the influence range changes drastically, then the rewinding is thorough and DyCLA explores the optimal seeds from scratch. The first case demonstrates the efficiency of DyCLA under the docile variations of the network structure, while the second case reflects the flexibility of DyCLA to discover a completely different subset of seeds.

The contributions of this paper are:
\begin{enumerate}
\item We apply the learning automata theory to the problem of influence maximization in dynamic social networks and propose Dynamic Conjugate Learning Automata.
\item The proposed method is apt in reflecting the continuous dynamics of the network, and is comparatively efficient.
\item Experimental results on both synthetic and real-world datasets verify the efficacy of the proposed method.
\end{enumerate}

This paper proceeds as follows: Section \ref{section:2} reviews the formulation of IM and some attempts to address IM in dynamic networks. Section \ref{section:3} presents the proposed method DyCLA. Section \ref{section:4} is devoted to experimental results and subsequent discussions. Section \ref{section:5} concludes the paper.

\section{Related Works}
\label{section:2}
People have long been studied how the decisions of people are affected by their neighbors and friends or how the "word-of-mouth" affects people's behavior.
Specifically, researchers have been paying attention to the diffusion processes of information among participants in social networks.
Various general diffusion models have been proposed, including the most basic independent cascade and linear threshold model \cite{ref:IM}. Motivated by marketing and advertising, Domingos and Richardson proposed the fundamental algorithmic problem of IM \cite{ref:rw_1}\cite{ref:rw_2} which aims to choose the few key individuals in a social network as the source nodes, or seeds, to maximize the spread of influence, i.e., the number of influenced nodes.

\textbf{Problem 1. Information Maximization.}
Given a network $\mathcal{G}$ and the number of seeds $K$, an information maximization algorithm aims to identify the optimal seed set $\mathcal{S}^{*}$ such
\begin{equation}
  \mathcal{S}^{*} = \arg\max_{\mathcal{S}\subseteq\mathcal{V},|\mathcal{S}|=K} \sigma(\mathcal{S})
,\end{equation}
where $\sigma(\mathcal{S})$ is the number of nodes influenced by seed set $\mathcal{S}$ and is defined according to some diffusion model.

The problem is essentially a combinatorial optimization problem and is NP-hard \cite{ref:IM}.
The naive greedy method is the simplest approach, it chooses the estimated optimal node one by one using Monte-Carlo simulations.
The ratio between the outcome of the greedy method and the optimal one is lower bounded by $(1-\frac{1}{e})$ \cite{ref:IM}.
Following these groundbreaking works, abundant greedy-based strategies have been proposed to improve the efficiency in choosing the seeds.
Among them, the Cost-Effective Lazy Forward algorithm (CELF) \cite{ref:celf} that takes advantage of the submodularity of the problem to compute marginal influential gain is the state-of-the-art and is 700 times faster than the naive greedy method.
Apart from greedy algorithms, the heuristic methods that explore the properties of $\mathcal{G}$ rather than taking $\sigma(\cdot)$ as a \emph{black box} further improve the efficiency of seed set choosing by narrowing the candidate set. Examples included degree-based methods and community-based methods \cite{ref:hm}.
However, the structures of social networks are often intractable in reality, and the outcome of a heuristic method can be arbitrarily bad due to the lack of a theoretical boundary.

LA, an elementary variant of reinforcement learning, is known for its adaptivity in a stochastic environment.
Therefore it has been successfully applied to solve the information maximization problem since $\sigma(\cdot)$ is a random function.
It turns out that the LA-based approach can be even faster than traditional greedy-based methods such as CELF when a single LA is utilized as an appropriate optimizer to find the seeds following the greedy paradigm \cite{ref:IMLA}.
The CLA based method is also proposed in the IM problem to obtain better-than-greedy results while preserving some degree of efficiency \cite{ref:IMCLA}.

The methods discussed above all operate in a static social network where the network structure is invariant.
For a dynamic social network that better depicts the reality, $\mathcal{G}$ itself varies between different network snapshots, which brings new challenges.

\textbf{Problem 2. IM in Dynamic Social Networks.}
For a series of snapshots of a network in different time: $\mathcal{G}_{0},\mathcal{G}_{1},\mathcal{G}_{2},\cdots,\mathcal{G}_{T}$ and the budget of seed set $K$, IM in dynamic social networks aims to maximize the snapshot-wise spread of influence, i.e., the total number of influenced nodes in all snapshots by choosing the time-dependent optimal set of seeds $\mathcal{S}^{*}(t)$ in each snapshot $t (t=0,1,2,\cdots,T)$ such that
\begin{equation}
    \mathcal{S}^{*}(t) = \arg\max_{\mathcal{S}(t)\subseteq\mathcal{V}(t),|\mathcal{S}(t)|=K} \sigma_{t}(\mathcal{S}(t)).
\end{equation}

The methods designed for static social networks have to learn the new set of seeds for each snapshot independently, even if the difference between snapshots turns out to be trivial. This inter-snapshot similarity leaves adequate space for improvement.
Zhuang \cite{ref:dy1} proposed an approximate method Maximum Gap Probing to track the change of dynamic networks through probing a small portion of the network.
Inspired by Zhuang, Han \cite{ref:dy2} improved the efficiency by probing communities instead of nodes.
However, both methods in \cite{ref:dy1} and \cite{ref:dy2} have the following limitations: (1). They both consider local topology which could turn out to be unreliable. (2). They are both degree-based (and thus heuristic) methods in which the nodes with top $K$ highest degree are chosen as the seeds, which might be ineffective.

In this paper, we generalize LA-based IM algorithms to dynamic networks and provide a new solution to \textbf{Problem 2}.

\section{Proposed Method}
\label{section:3}
\subsection{Learning Automaton}
As an elementary paradigm in reinforcement learning, an LA adaptively explores the optimal action that maximizes the reward among all possible choices by interacting with a stochastic environment.
An LA with its environment is formalized as a triplet $<\mathcal{A}, \mathcal{B}, \mathbf{D}>$, where $\mathcal{A} = \{\alpha_{1}, \alpha_{2},\cdots\}$ is the set of possible actions, $\mathcal{B} = \{\beta_{1},\beta_{2},\cdots\}$ is the set of possible feedback from the environment, and $\mathbf{D}$ is the reward matrix of the environment following

\begin{equation}
\label{eq:interacting}
  \Pr\{\beta_{q}|\alpha_{r}\} = d_{r,q}, \beta_{q}\in \mathcal{B},\alpha_{r}\in \mathcal{A}.
\end{equation}

When $\mathbf{D}$ is fixed, the environment is called \emph{stable}, otherwise, it is an \emph{unstable} environment. For most LA schemes, the training is equivalent to tuning the normalized action probability vector $\mathbf{P} = [\mathbf{P}_{1},\mathbf{P}_{2},\cdots]$ to maximize the expected reward $\sum_{r,q} d_{r,q}\cdot \mathbf{P}_{r}\cdot\beta_{q}$.
The training process consists of a number of iterations, during the $i$-th iteration, the LA selects the action $\alpha(i)$ according to $\mathbf{P}(i)$
\begin{equation}
\label{eq:selection}
  \Pr\{\alpha(t) = \alpha_{r}\} = \mathbf{P}_{r}(i).
\end{equation}

The environment receives $\alpha(i)$ and returns the feedback $\beta(i)$ satisfying \eqref{eq:interacting}. The LA receives $\beta(i)$ and updates $\mathbf{P}(i)$ into $\mathbf{P}(i+1)$ according to some specific strategy, this is often done jointly with some additional information especially estimators \cite{ref:LA} denoted by $\mathbf{E}$, formally
\begin{equation}\label{eq:PE}
\mathbf{P}(i+1),\mathbf{E}(i+1)=\mathcal{L}\left(\mathbf{P}(i),\mathbf{E}(i),\alpha(i),\beta(i)\right).
\end{equation}
The internal state of an LA constitutes of both $\mathbf{P}$ and $\mathbf{E}$. An LA gets converged and terminates its training when $\max_{r}\left\{\mathbf{P}_{r} \right\}>\mathcal{T}$, where $\mathcal{T}$ is a predefined threshold.

\subsection{Conjugate Learning Automata in Influence Maximization}
\begin{figure}[htbp]
\centering
\begin{tikzpicture}[scale = 0.25]
\draw [black] (0,0) circle [radius = 1];
\draw [black] (0,6) circle [radius = 1];
\draw [black] (0,9) circle [radius = 1];
\node at (0,0) [scale = 0.5] {$\text{LA}^{K}$};
\node at (0,6) [scale = 0.5]{$\text{LA}^{2}$};
\node at (0,9) [scale = 0.5]{$\text{LA}^{1}$};
\draw (1,0)--(3,0) [-latex];
\draw (1,6)--(3,6) [-latex];
\draw (1,9)--(3,9) [-latex];
\node at (4.5,0) [scale = 0.75] {$\alpha^{K}(i)$};
\node at (4.5,6) [scale = 0.75] {$\alpha^{2}(i)$};
\node at (4.5,9) [scale = 0.75] {$\alpha^{1}(i)$};
\node at (2,3) {$\cdots\cdots$};
\draw[decorate,decoration={calligraphic brace,mirror,amplitude=3mm,}][thick] (6,0) -- (6,9);
\node at (9.5,4.5) {$\alpha^{\text{CLA}}(i)$};
\draw (12,4.5)--(13.5,4.5) [-];
\draw (13.5,4.5)--(13.5,-3) [-];
\draw (13.5,-3)--(3,-3) [-latex];
\draw (-3,-2) rectangle (3,-4);
\node at (0, -3) {Network};
\draw (-3,-3)--(-10,-3) [-];
\draw (-10,-3)--(-10,3) [-latex];
\node at (-10,4.5) {$\beta(t)$};
\draw (-8.5,4.5)--(-2.5,4.5) [-latex];
\draw[decorate,decoration={calligraphic brace,amplitude=3mm,}][thick] (-1.3,0) -- (-1.3,9);
\node at (-6.25, 5.5) [scale = 0.75] {Update};
\draw [red,dashed,rounded corners] [thick] (-3.5,7.8) rectangle (3.5,-5);
\end{tikzpicture}
\caption{CLA with $K$ LAs. The external environment of $\text{LA}^{1}$ is besieged by the dashed red line.}
\label{figure:CLA}
\end{figure}
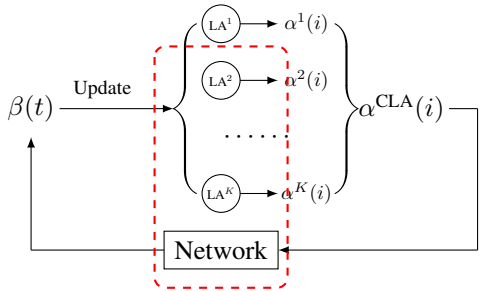
The conjugate learning automata (CLA) follows the paradigm of automata game \cite{ref:LAG} and is capable of avoiding adverse pitfalls that confine the performance of ordinary greedy algorithms in IM \cite{ref:IMCLA}. In the context of IM, the CLA consists of the following elements: (where we use superscript to denote the index of seeds and subscript to that of vertices)
\begin{itemize}
\item{\emph{$K$ individual LAs:} $\text{LA}^{1},\text{LA}^{2},\cdots,\text{LA}^{K}$ (where $K$ is the number of seeds), each of which select one seed, thus $\mathcal{A}=\mathcal{V}^{K}$. We illustrate a CLA comprising $K$ individual LAs in Figure. \ref{figure:CLA}.}

\item{\emph{Action selection:}} At the $i$-th iteration, an individual learning automaton $\text{LA}^{k}$ selects an action $\alpha^{k}(i)\in \mathcal{V}$ according to its probability vector $\mathbf{P}^{k}(i)$. These individual choices are then connected as $\alpha^{\text{CLA}}(i)=\mathcal{S}(t)=\left\{\alpha^{1}(i),\alpha^{2}(i),\cdots,\alpha^{K}(i)\right\}$, the action of the CLA.

\item{\emph{Environment:} The stochastic environment of CLA is the stochastic propagation function $\sigma(\cdot)$ and $\mathcal{B}=\mathbb{N}$, $\mathbf{D}$ is determined by both the network structure and the set of seeds. A response $\beta(i)=\sigma(\alpha^{\text{CLA}}(i))$ is returned.}

\item{\emph{Learning scheme:} The environment w.r.t. an individual $\text{LA}^{k}$ consists of both the static network and other $(K-1)$ LAs, and could vary during the learning procedure. To ensure the convergence of $\text{LA}^{k}$, when $\text{LA}^{k}$ is updating its internal states, all the rest $(K-1)$ LAs are kept fixed. Meanwhile, it is undesirable that only one LA converges while others are left totally untrained. So we set a temporary halt threshold $\delta < \mathcal{T}$. The learning of CLA consists of a number of rounds, at each round, each of the $K$ LAs undergoes training while the probability vectors of other $(K-1)$ LAs are fixed. Training only one LA during one round is no different from training an LA in a stable environment, only with the halt condition substituted by $\max_{r}\left\{\mathbf{P}_{r}\right\}>\delta$. After one round terminated, we increase the value of $\delta$ and start another round. The procedure is finished when $\delta \geq \mathcal{T}$. This scheme is summarized as in Algorithm. \ref{Alg:ICLA}.}
\end{itemize}

\begin{algorithm}
  \caption{CLA for IM}
  \label{Alg:ICLA}
  \begin{algorithmic}[1]
  \STATE \textbf{Input} Initial Temporary threshold: $\delta=\delta_{0} \in (0,1)$.
  \STATE \textbf{Input} Iterative increment: $\Delta\delta \in (0,1)$.
  \STATE \textbf{Input} Convergence threshold: $\mathcal{T}\in (0,1)$.
  \STATE \textbf{Initialize} Convergence flag: $\mathcal{I} = 1$.
  \STATE \textbf{Initialize} $K$ LAs.
  \STATE \textbf{Initialize} $i=0$
  \REPEAT
  \FOR {$k=1$ to $K$}
  \REPEAT
  \STATE $\alpha^{\text{CLA}}(i)=\left\{\alpha^{1}(i),\alpha^{2}(i),\cdots,\alpha^{K}(i)\right\}$, s.t.\eqref{eq:selection}.
  \STATE $\beta(i)=\sigma(\alpha^{\text{CLA}}(i))$.
  \STATE \textbf{Update} $\mathbf{P}^{k}(i+1),\mathbf{E}^{k}(i+1)$ using \eqref{eq:PE}.
  \STATE $\mathbf{P}^{s}(i+1),\mathbf{E}^{s}(i+1)=\mathbf{P}^{s}(i),\mathbf{E}^{s}(i),\forall s\neq k.$
  \STATE $++i.$
  \UNTIL {$\max_{n}\{\mathbf{P}^{k}_{n}(i)\} \geq \delta$}.
  \ENDFOR
  \IF {$\delta < \mathcal{T}$}
  \STATE $\delta = \min\left\{\delta+\Delta\delta,\mathcal{T}\right\}$.
  \ELSE
  \STATE $\mathcal{I} = 0$.
  \ENDIF
  \UNTIL $\mathcal{I} = 0$.
  \STATE \textbf{Output} $\mathcal{S}=\left\{v_{\arg\max_{n=1}^{N}\left\{\mathbf{P}^{k}_{n}(i) \right\}} \right\}_{k=1}^{K}$.
  \end{algorithmic}
\end{algorithm}

Having formulated the update framework of CLA, we specify the update strategy of one individual LA, namely line 12 in Algorithm. \ref{Alg:ICLA}.
We extend the classic LA algorithm, discretized general pursuit algorithm (DGPA) \cite{ref:DGPA} which enjoys concise structure, low computation cost and analytic optimality. The extended version (eDGPA) for the $k$-th individual LA is summarized in Algorithm. \ref{Alg:DGPA}, with $\mathbf{E}=\left(\mathbf{Z},\mathbf{R}\right)$.

\begin{algorithm}
  \caption{The extended version of DGPA, eDGPA as $\mathcal{L}$}
  \label{Alg:DGPA}
  \begin{algorithmic}[1]
  \STATE \textbf{Initialize} Update Step Size: $\Delta = \frac{1}{\mathcal{R}\cdot N}$, where $\mathcal{R}$ is the resolution parameter.
  \STATE \textbf{Initialize} Action probability vector: $\mathbf{P}^{k}(0) = \frac{1}{N}\textbf{1}_{N}$.
  \STATE \textbf{Initialize} Action selection times vector: $\mathbf{Z}^{k}(0) = \textbf{1}_{N}$.
  \STATE \textbf{Initialize} Rewards estimation vector: $\mathbf{R}^{k}(0) = \textbf{1}_{N}$.

  \STATE \textbf{Input }$i,\alpha(i),\beta(i)$:
  \STATE \quad Suppose $\alpha(i) = v_{n}$.
  \STATE \quad \textbf{Update} $\mathbf{R}^{k}_{n}(i+1) = \frac{\mathbf{Z}^{k}_{n}(i)\cdot \mathbf{R}^{k}_{n}(i) + \beta(i)}{\mathbf{Z}^{k}_{n}(i)+1},$
  \STATE \quad \textbf{Update} $\mathbf{R}^{k}_{s}(i+1)=\mathbf{R}^{k}_{s}(i), \forall s=1,2,\cdots,N,s\neq n.$
  \STATE \quad \textbf{Update} $\mathbf{Z}^{k}_{n}(i+1)=\mathbf{Z}^{k}_{n}(i)+1.$
  \STATE \quad \textbf{Update} $\mathbf{Z}^{k}_{s}(i+1)=\mathbf{Z}^{k}_{s}(i),\forall s\neq n$
  \STATE \quad $W_{n}(i)=|\left\{s:s=1,\cdots,N, \mathbf{R}^{k}_{s}(i+1)> \mathbf{R}^{k}_{n}(i+1) \right\}|$.
  \STATE
    \quad \textbf{Update} $\mathbf{P}^{k}_{s}(i+1) = \min\{\mathbf{P}^{k}_{s}(i) + \frac{\Delta}{W_{n}(i)}, 1\},\forall s=1,2,\cdots,N: \mathbf{R}^{k}_{s}(i+1)> \mathbf{R}^{k}_{n}(i+1)$,
  \STATE
    \quad \textbf{Update} $\mathbf{P}^{k}_{s}(i+1) = \max\{\mathbf{P}^{k}_{s}(i) - \frac{\Delta}{N-W_{n}(i)}, 0\},\forall s=1,2,\cdots,N: \mathbf{R}^{k}_{s}(i+1)<\mathbf{R}^{k}_{n}(i+1)$,
  \STATE
    \quad \textbf{Update} $\mathbf{P}^{k}_{n}(i+1) = 1-\sum\limits_{s=1,s\neq n}^{N} \mathbf{P}^{k}_{s}(i+1)$.

  \end{algorithmic}
\end{algorithm}

\subsection{Conjugate Learning Automata for Dynamic Networks}
Before formalizing Dynamic CLA that adapts to dynamic networks, we shall consider two extreme cases:
\begin{itemize}
\item \emph{Docile variation}: If the influence range of the current choice of seeds is not dramatically affected, then the algorithm should only make a slight revision instead of restarting from the beginning. In this case the historical knowledge can be partially trusted and taken advantage of. As Figure. \ref{figure:2} ($t=1$).
\item \emph{Drastic variation}: If the influence range of the current set of seeds is rapidly increased/decreased, then the algorithm should erase a larger portion of (potentially all) memory and restart itself on this probably new network. In this manner, the algorithm can cope with some peculiar structure variations that potentially bring significant change to the network behavior. As Figure. \ref{figure:2} ($t=2$).
\end{itemize}

\begin{figure}[htbp]
\centering
\includegraphics[width = 0.48\textwidth]{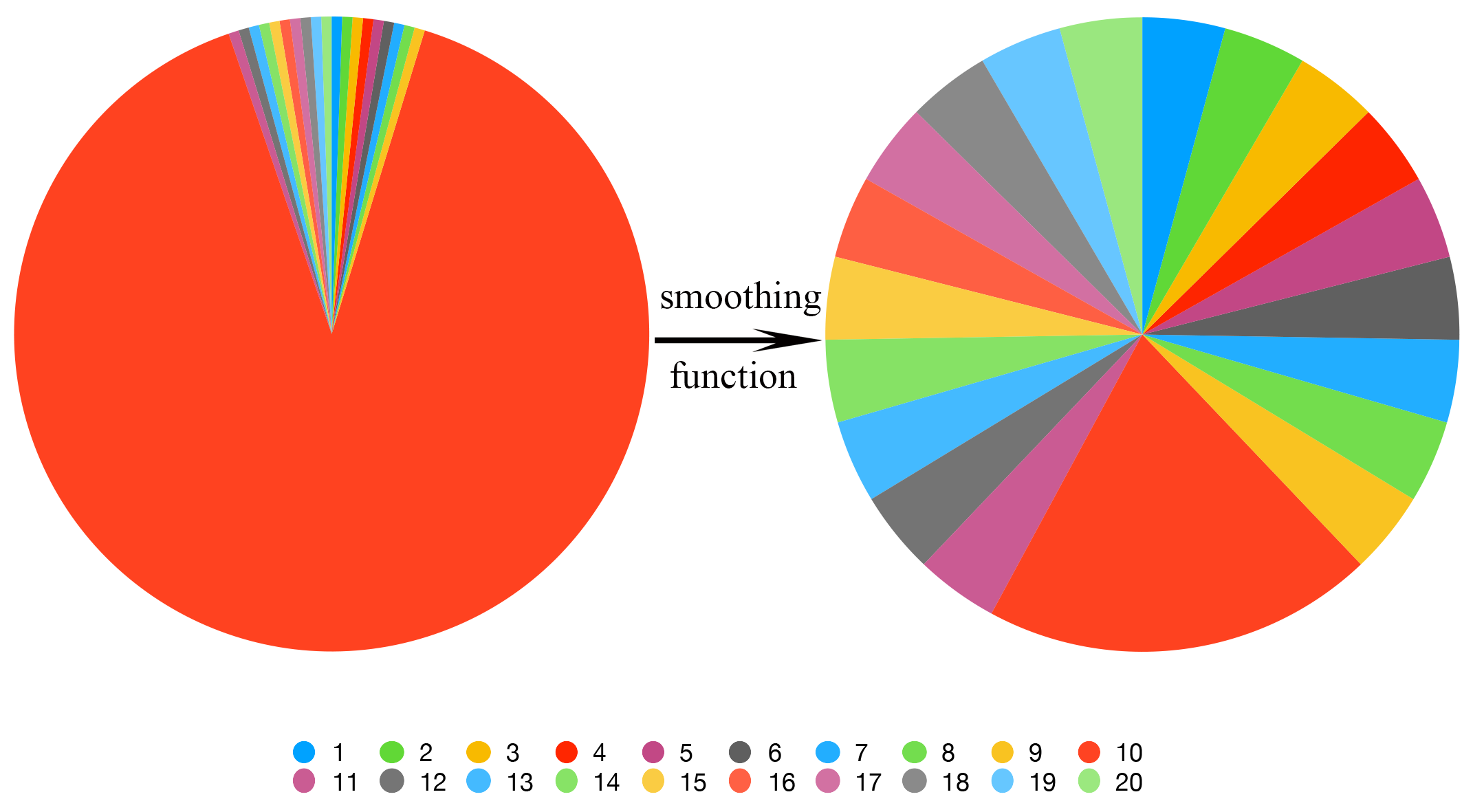}
\caption{The visualization of $\mathbf{P}$ being smoothed by the smoothing function, $N=20$.}
\label{figure:Smoothing}
\end{figure}

To conclude, the variation of the network structure results in the variation of the influence range, and should be reflected by the variation of the algorithm' s estimation of the current network. Since CLA has a collection of probability vectors and estimators as a model of memory, it is straightforward to incorporate the observations from the two cases above into CLA:
\begin{enumerate}
\item Firstly, after the variation of the structure takes place, the propagation range of the current set of seeds $\mathcal{S}$ becomes $\sigma'(\mathcal{S})$. The significance of this variation is measured by the difference in influence range
\begin{equation}
\Delta \sigma =|\sigma'(\mathcal{S}) -\sigma(\mathcal{S})|.
\end{equation}
\item Secondly, the convergence of any individual LA in the CLA is relaxed by a parameterized \emph{smoothing function} $f({\Delta\sigma},\cdot)$ that maps an $N$-dimensional simplex to another $N$-dimensional simplex, during which the maximal component of the input is reduced. The larger $\Delta\sigma$ is, the more closely the output turns to be a uniform distribution. Essentially, $f({\Delta\sigma},\cdot)$ is the inverse of the update process of an LA. A small $\Delta \sigma$ cancels only the influence of the latest few rounds in Algorithm. \ref{Alg:ICLA}, and CLA should be able to find another optimal solution quickly. A large $\Delta \sigma$ cancels almost all information as if CLA has just been initialized.
Figure. \ref{figure:Smoothing} visualizes how the smoothing function acts upon the action probability vector $\mathbf{P}$ of a specific LA. Where the portion of the $n$-th component denotes the corresponding probability $\mathbf{P}_{n}$.
\item Thirdly, the information in the estimator has to be perturbed. Any components that record the expected propagation range of a choice is added with a zero-mean stochastic \emph{perturbation} $\Delta R(\Delta\sigma)$ whose variance is monotonic with $\Delta \sigma$. Meanwhile, the selection times vector is set to be noninformative. In this manner the individual LAs are encouraged to explore new candidate seeds.
\end{enumerate}

Formally, we adopt the following choice of \emph{smoothing function} and $\emph{perturbation}$
\begin{equation}
f_{\phi}(\Delta \sigma, \mathbf{P}):
\begin{cases}
  \mathbf{P}_{s}=\mathbf{P}_{s}+\frac{1}{R-1}\cdot \psi, \forall s=1,2,\cdots,N, s\neq m
  \\\mathbf{P}_{m} = \mathbf{P}_{m}-\psi
\end{cases}
,\end{equation}
where $m = \arg\max_{n}\{\mathbf{P}_{n}\}$ and
 \[\psi =\min\left\{ \phi\cdot\frac{\Delta\sigma}{\sigma'(\mathcal{S})+\sigma(\mathcal{S})}\cdot\mathbf{P}_{m},1\right\}.\]

\begin{equation}
\Delta R(\Delta \sigma) \sim \mathcal{N}(0,\Delta \sigma).
\end{equation}
At last, the DyCLA can be formulated as Algorithm. \ref{Alg:dyCLA}, where $\texttt{CLA\_for\_IM}$ is Algorithm. \ref{Alg:ICLA} with line 6-7 cancelled and $\sigma(\cdot)$ replaced by $\sigma_{t}(\cdot)$.
\begin{algorithm}
  \caption{Dynamic CLA}
  \label{Alg:dyCLA}
  \begin{algorithmic}[1]
  \STATE \textbf{Input} Initial Temporary threshold: $\delta_{0} \in (0,1)$.
  \STATE \textbf{Input} Iterative increment: $\Delta\delta \in (0,1)$.
  \STATE \textbf{Input} Convergence threshold: $\mathcal{T}\in (0,1)$.
  \STATE \textbf{Input} Smoothins parameter: $\phi$.
  \STATE \textbf{Initialize} $K$ LAs, $i=0,t=0$.
  \STATE \textbf{Output}$\mathcal{S}(t)=\texttt{CLA\_for\_IM}(\delta_{0},\Delta \delta,\mathcal{T}, i, t)$.
  \REPEAT
  \STATE $R(t)=\sigma_{t}(\mathcal{S}(t))$.
  \STATE $++t.$\quad //Network structure changes.
  \STATE $\Delta\sigma(t) = |\sigma_{t}(\mathcal{S}(t-1))-R(t)|$.
  \FOR {$k=1$ to $K$}
  \STATE $\mathbf{P}^{k}=f_{\phi}(\Delta\sigma(t),\mathbf{P}^{k})$.
  \STATE $\mathbf{Z}^{k}=\frac{\mathbf{1}_{N}^{\text{T}}\mathbf{Z}^{k}}{N}\cdot\mathbf{1}_{N}$.
  \FOR {$n=1$ to $N$}
  \STATE $\Delta R \sim \mathcal{N}(0,\Delta\sigma(t))$.
  \STATE $\mathbf{R}^{k}_{n}+=\Delta R$.
  \ENDFOR
  \ENDFOR
  \STATE \textbf{Output} $\mathcal{S}(t)=\texttt{CLA\_for\_IM}(\delta_{0},\Delta \delta,\mathcal{T},i,t)$.
  \UNTIL No more snapshots.
  \end{algorithmic}
\end{algorithm}

The advantages of the proposed Dynamic CLA method over established methods are as follows:
\begin{itemize}
\item {The significance of a network structure variation is measured only through the corresponding difference in the influence range. For large networks, it is generally unfair to evaluate the impact of a structure variation only through local topology as in \cite{ref:dy1}\cite{ref:dy2}.}
\item{DyCLA does not explicitly distinguish a static network from dynamic ones (let $\Delta\sigma=0$ in line 9 in Algorithm. \ref{Alg:dyCLA} then line 10-16 keep the CLA intact and line 17 makes no difference to original $\mathcal{S}$). Therefore our proposal degenerates gracefully to its counterpart for static networks, while for some proposals, the version for static networks and that for dynamic ones are differentiated.}
\item{For a fixed snapshot of the network, our proposal yields a relatively sophisticated choice of $\mathcal{S}$. Due to the inconsistency between static and dynamic cases and the demand of efficiency, many proposals have cut down the resource devoted to a fixed snapshot in time series. Meanwhile, DyCLA with good consistency and high efficiency can preserve the same efficacy compared with algorithms that are specialized for static networks.}
\end{itemize}

\section{Experimental Results and Discussions}
\label{section:4}
\subsection{Experimental Settings}
All experiments are conducted under the popular weighted cascade model \cite{ref:hm}, where a node $v \in \mathcal{V}$ is influenced by edge $(u,v)\in\mathcal{E}$ with probability $\frac{1}{\text{in-degree}(v)}$ in directed graphs or $\frac{1}{\text{degree}(v)}$ in undirected graphs.
For comparison, the representative greedy-based algorithm CELF \cite{ref:celf} and the greedy LA-based algorithm IMLA \cite{ref:IMLA} serve as the baselines.
The heuristic methods have generally smaller spread range and are saved from comparisons.
For parameters in DyCLA, the resolution parameter $\mathcal{R}$ for eDGPA  is set as $K$.
($\delta_{0},\Delta\delta,\mathcal{T}$) are parameterized as ($\frac{1}{K},\frac{1}{2K},0.999$) in all experiments, which is the same as \cite{ref:IMCLA}.
For CELF, the number of Monte-Carlo simulations is uniformly set to 10000, following the consensus in literature.

\subsection{Toy Example}

\begin{figure*}[htbp]
\centering
\includegraphics[width = \textwidth]{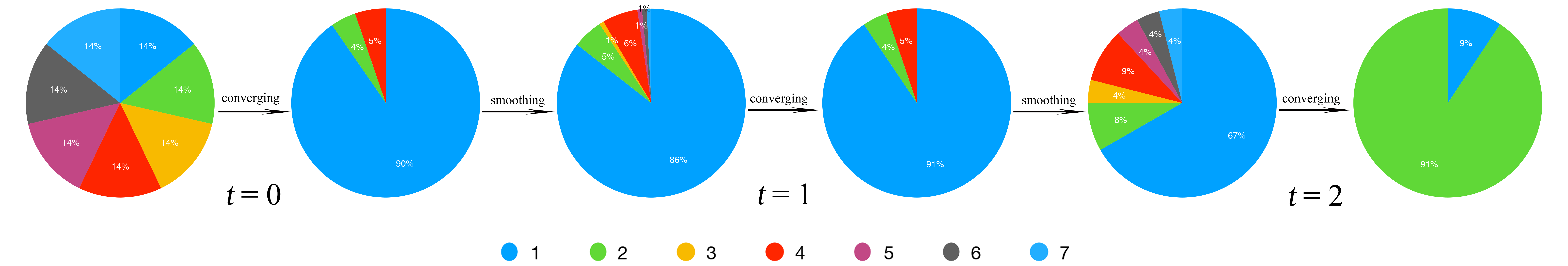}
\caption{The changes of $\mathbf{P}$ during the IM process among snapshots.}
\label{figure:5}
\end{figure*}

Figure. \ref{figure:5} visualizes the change of $\mathbf{P}$ across snapshots in Figure. \ref{figure:2}.
It can be observed that: (1). DyCLA correctly converges to the optimum at all snapshots. (2). When the variation of the influence is docile ($t=1$), the smoothing function helps to memorize useful knowledge and accelerate convergence. (3). When the influence range of outdated seeds is significantly declined, DyCLA can veto the previous decision and converge to a better choice ($t=2$).

\subsection{Large-Scale Networks}
\begin{figure}[htbp]
\centering
\includegraphics[width = 0.48\textwidth]{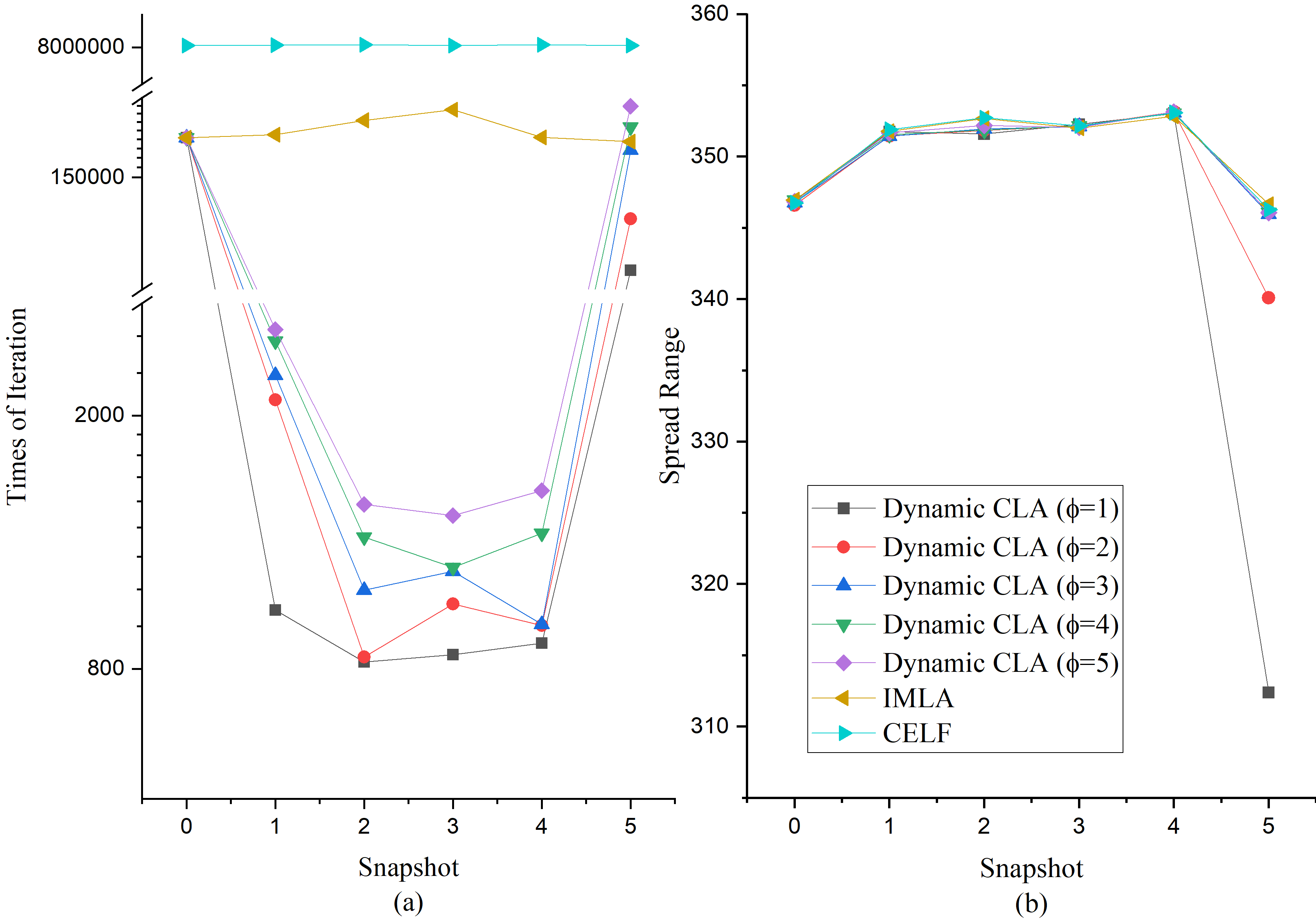}
\caption{The required number of interactions with the network and the spread range of different algorithms in synthetic network with $N=800,K=5$.}
\label{figure:LargeScale}
\end{figure}

\begin{figure}[htbp]
\centering
\includegraphics[width = 0.48\textwidth]{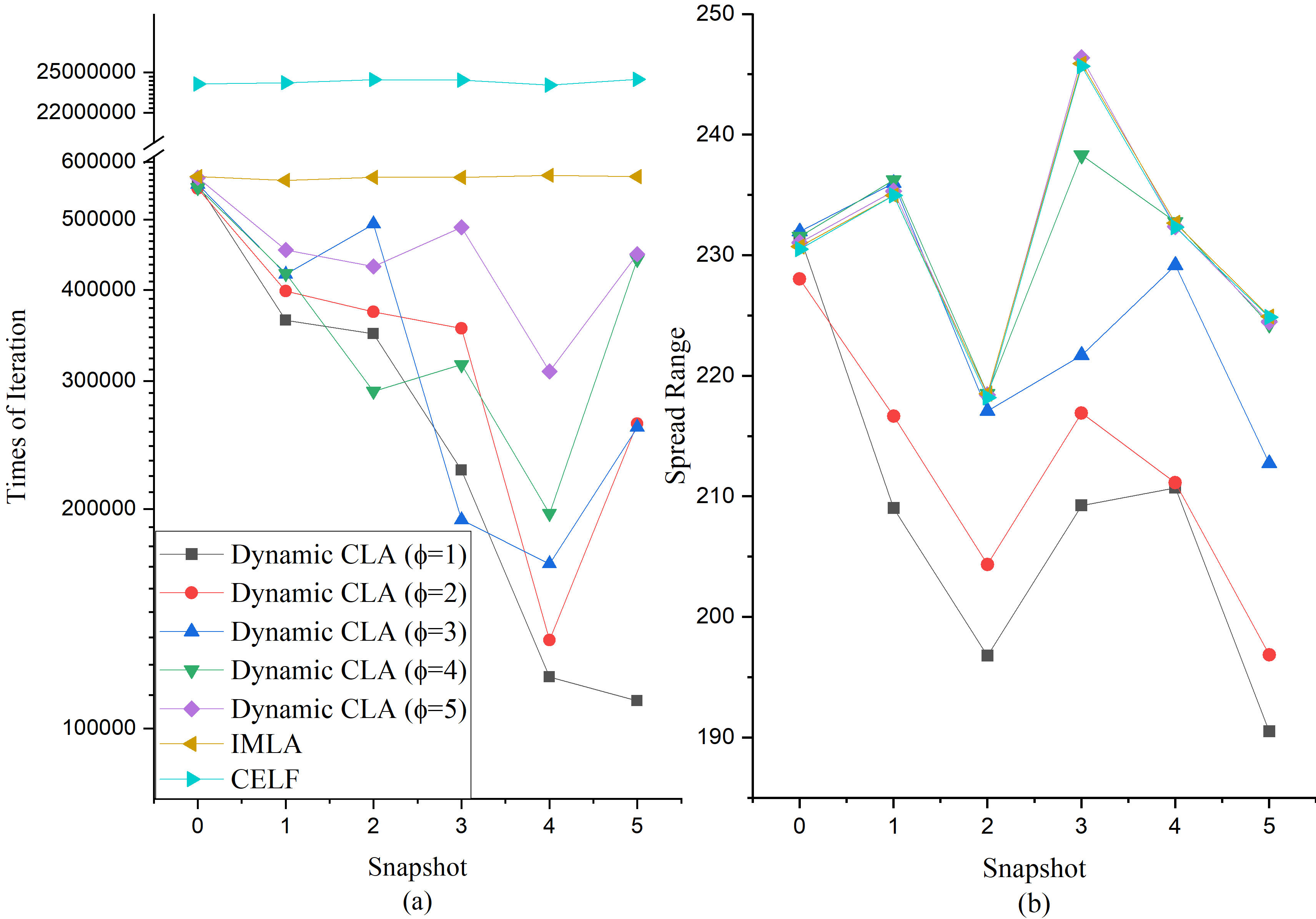}
\caption{The required number of interactions with the network and the spread range of different algorithms in real-world network with $N=2359,K=5$.}
\label{figure:RealWorld}
\end{figure}

Further evaluations on large-scale networks are conducted on both synthetic datasets and real-world datasets.
The metrics of interest are the required number of interactions with the network, i.e. the times we calculate $\sigma(\cdot)$, and the spread range of different algorithms in different snapshots.
The synthetic dataset with $800$ vertices and $6$ snapshots is randomly generated.
The structure variation from $t=0$ to $t=4$ are docile (the optimal set of seeds are identical), while a drastic variation is coined before $t=5$.
For real-world dataset, we adopt Enron dataset \cite{ref:Enron} which collects e-mail interconnections from 150 senior executives, altogether 2359 users are involved.
We use the record from December 1999 to April 2000 as $6$ snapshots (each snapshot corresponds to one month) that constitute the real-world dynamic social network dataset. In both cases we let $K=5$.
Figure. \ref{figure:LargeScale} and Figure. \ref{figure:RealWorld} present the experimental results, from which we can observe:
\begin{itemize}
\item DyCLA converges quickly after docile variations and thousands of times of interactions are reduced compared with algorithms for static networks, this fact verifies its effectiveness, as shown in Figure. \ref{figure:LargeScale}(a) and Figure. \ref{figure:RealWorld}(a).
\item Figure. \ref{figure:LargeScale}(a) and Figure. \ref{figure:RealWorld}(a) indicates that smaller $\phi$ leads to higher efficiency, \emph{vice versa}. This is in accordance with the intuition since smaller $\phi$ means less cancellation of the memory and faster convergence.
\item Figure. \ref{figure:LargeScale}(b) and Figure. \ref{figure:RealWorld}(b) shows that if $\phi$ is too small then DyCLA might fail to discover the optimal seeds, while a larger $\phi$ ensures correct convergence. Intuitively, if DyCLA erases little knowledge, then it would overfit the previous network and be trapped in a local optimum. Therefore $\phi$ reflects the trade-off between accuracy and efficiency in DyCLA.
\end{itemize}

\section{Conclusion}
\label{section:5}
In this paper we propose DyCLA to address the problem of influence maximization in dynamic social networks. By incorporating the inverse of the update procedure, DyCLA straightforwardly generalizes its counterpart for static social networks. This generalization comprehensibly reflects our intuition about the structure variation. Experimental results verify the privileges of DyCLA against established methods.

\section*{Acknowledgment}
This work was supported by the National Nature Science Foundation of China under Grant 61971283.



\begin{thebibliography}{00}
  \bibitem{ref:IM}D. Kempe, J. Kleinberg, and E. Tardos, "Maximizing the spread of influence through a social network," in \emph{Proceedings of the ninth ACM
  SIGKDD international conference on Knowledge discovery and data
  mining.} ACM, 2003, pp. 137–146.
  \bibitem{ref:App1}C Wei, C Wang, and Y Wang. "Scalable influence maximization for prevalent viral marketing in large-scale social networks." in \emph{Proceedings of the 16th ACM SIGKDD international conference on Knowledge discovery and data mining.} ACM, 2010, pp.1029-1038.
  \bibitem{ref:App2}Leskovec, J., Krause, A., Guestrin, C., Faloutsos, C., VanBriesen, J., and Glance, N. "Cost-effective outbreak detection in networks" in \emph{Proceedings of the 13th ACM SIGKDD international conference on Knowledge discovery and data mining.} ACM, 2007, pp.420-429.
  \bibitem{ref:dy1}H Zhuang, Y Sun, J Tang, J Zhang, and X Sun. "Influence maximization in dynamic social networks." \emph{2013 IEEE 13th International Conference on Data Mining.} IEEE, 2013.
  \bibitem{ref:dy2}M Han, M Yan, Z Cai, Y Li, X Cai, and J Yu. "Influence maximization by probing partial communities in dynamic online social networks." \emph{Transactions on Emerging Telecommunications Technologies} 28.4 (2017): e3054.
  \bibitem{ref:dy3}Y Bao, X Wang, Z Wang, C Wu, and Francis C.M. Lau. "Online influence maximization in non-stationary social networks." \emph{2016 IEEE/ACM 24th International Symposium on Quality of Service (IWQoS)}. IEEE, 2016.
  \bibitem{ref:IMLA}H Ge, J Huang, C Di, J Li, and S Li. "Learning automata based approach for influence maximization problem on social networks." \emph{IEEE Second International Conference on Data Science in Cyberspace (DSC).} IEEE, 2017, pp.108-117.
  \bibitem{ref:IMCLA}C Di, F Li, K Qi, and S Li. "Maximizing Influence on Social Networks with Conjugate Learning Automata." \emph{2019 IEEE Global Communications Conference}. IEEE, 2019.
  \bibitem{ref:rw_1}P. Domingos and M. Richardson, "Mining the network value of customers," in \emph{Proceedings of the seventh ACM SIGKDD international conference on Knowledge discovery and data mining.} ACM, 2001, pp. 57–66.
  \bibitem{ref:rw_2}Richardson, Matthew, and Pedro Domingos. "Mining knowledge-sharing sites for viral marketing." \emph{Proceedings of the eighth ACM SIGKDD international conference on Knowledge discovery and data mining}. ACM, 2002.
  \bibitem{ref:celf}Leskovec, J., Krause, A., Guestrin, C., Faloutsos, C., VanBriesen, J., and Glance, N. "Cost-effective outbreak detection in networks" in \emph{Proceedings of the 13th ACM SIGKDD international conference on Knowledge discovery and data mining.} ACM, 2007, pp.420-429.
  \bibitem{ref:hm}Y Li, J Fan, Y Wang, and KL Tan. "Influence maximization on social graphs: A survey." \emph{IEEE Transactions on Knowledge and Data Engineering} 30.10 (2018): 1852-1872.
  \bibitem{ref:LA}Narendra, Kumpati S., and Mandayam AL Thathachar. \emph{Learning automata: an introduction}. Courier Corporation, 2012.
  \bibitem{ref:LAG}Fu, King-Sun, and Timothy J. Li. "Formulation of learning automata and automata games." \emph{Information Sciences} vol.1, no.3, pp.237-256, 1969.
  \bibitem{ref:DGPA}Agache, Mariana, and B. John Oommen. "Generalized pursuit learning schemes: new families of continuous and discretized learning automata." \emph{IEEE Transactions on Systems, Man, and Cybernetics, Part B (Cybernetics)} vol.32, no.6, pp.738-749, 2002.
  \bibitem{ref:Enron}L Tang, H Liu, J Zhang, Z Nazeri. "Community evolution in dynamic multi-mode networks." \emph{Proceedings of the 14th ACM SIGKDD international conference on Knowledge discovery and data mining}. ACM, 2008.
\end{thebibliography}
\end{document}